\documentclass[3p,times,twocolumn]{elsarticle}
\biboptions{comma,sort&compress}
\usepackage{graphicx}
\usepackage[utf8]{inputenc}
\usepackage{here}
\usepackage{color}
\usepackage{ecrc}

\usepackage{tikz}
\usetikzlibrary{positioning,arrows}
\usetikzlibrary{decorations.pathmorphing}
\usetikzlibrary{decorations.markings}
\usepackage{pgfplots}
\usepackage{xparse}
\usetikzlibrary{positioning,arrows,patterns}
\usetikzlibrary{decorations.markings}
\usetikzlibrary{calc}

\volume{00}

\firstpage{1}

\journalname{Nuclear and Particle Physics Proceedings}

\runauth{}

\jid{nppp}

\jnltitlelogo{Nuclear and Particle Physics Proceedings}

\usepackage{amssymb}

\usepackage[figuresright]{rotating}

\def\bc{\begin{center}}
\def\ec{\end{center}}
\def\be{\begin{equation}}
\def\ee{\end{equation}}
\def\beqn{\begin{eqnarray}}
\def\eeqn{\end{eqnarray}}
\def\no{\nonumber}
\def\nn{\no\\}
\def\eqn#1{(\ref{#1})}
\def\ba{\begin{array}{c}}
\def\ea{\end{array}}
\def\bat{\begin{array}{cc}}
\def\bi{\begin{itemize}}
\def\ei{\end{itemize}}
\def\cA{{\cal A}}
\def\cL{{\cal L}}

\def\cO{{\cal O}}

\begin{document}

\begin{frontmatter}

\title{Nonleptonic $K\to 2\,\pi$ decay dynamics $^*$}

\cortext[cor0]{Talk given at 23th International Conference in Quantum Chromodynamics (QCD 20),  27 - 30 October 2020, Montpellier - FR}

\author{Hector Gisbert}
\ead{hector.gisbert@tu-dortmund.de}
\address{Fakultät für Physik, TU Dortmund, Otto-Hahn-Str.\,4, D-44221 Dortmund, Germany}

\pagestyle{myheadings}
\markright{ }
\begin{abstract}
Using Chiral Perturbation Theory to properly account for the dynamics of nonleptonic $K\to 2\,\pi$ decays, we found the Standard Model prediction for the CP violating ratio $\mbox{Re}\left(\varepsilon'/\varepsilon\right)=\left(14\pm 5\right) \times 10^{-4}$, where isospin breaking effects are included, in perfect agreement with the current experimental world average. Similar results have been reported by a recent release of improved lattice data.
\end{abstract}


\end{frontmatter}

\section{Introduction}

The large asymmetry between matter and antimatter in the observable universe requires the existence of additional CP sources beyond the Standard Model (SM). Accordingly, the CP violating ratio $\varepsilon'/\varepsilon$ provides an excellent road towards the discovery of New Physics (NP), since it is linearly proportional to the only existing CP source in the SM. In addition, its study offers a formidable test of flavor-changing neutral-currents (FCNCs) transitions and therefore the Glashow-Iliopoulos-Maiani (GIM) mechanism.

The current experimental world average of $\varepsilon'/\varepsilon$ from NA48~\cite{Batley:2002gn} and KTeV~\cite{AlaviHarati:2002ye,Abouzaid:2010ny},
\begin{equation}\label{eq:exp}
\hspace{1cm}\mathrm{Re}\, (\varepsilon'/\varepsilon)_{\,\mathrm{exp}}\; =\;
(16.6 \pm 2.3) \cdot  10^{-4}\, ,
\end{equation}
exhibits the existence of direct CP violation in decay transitions of $K\to 2\,\pi$, in addition, it is highly sensitive to new sources of CP violation due to its small size.

Since the establishment of its experimental measurement, the SM prediction of $\varepsilon'/\varepsilon$ has been subject of several years of discussions. The first next-to-leading order (NLO) calculations obtained SM values of order $\sim 10^{-4}$~\cite{Buras:1993dy,Buras:1996dq,Bosch:1999wr,Buras:2000qz,Ciuchini:1995cd,Ciuchini:1992tj}, in clear conflict with Eq.~\eqn{eq:exp}. However, it was soon realised by the authors of Refs.~\cite{Pallante:1999qf,Pallante:2000hk,Pallante:2001he}, that an important ingredient had been missed in the previous predictions, the inclusion of final-state interactions (FSI) between the two emitted pions. These corrections were computed within the low-energy effective realization of quantum chromodynamics (QCD), that is Chiral Perturbation theory ($\chi$PT). Once they were included in the calculation, the SM prediction was found consistent with Eq.~\eqn{eq:exp}, albeit with large uncertainties of non-perturbative origin. 

The development of advanced lattice QCD techniques as well as the increasing computer capabilities resulted in a successful calculation of the $\Delta I =3/2$  $K^+\to \pi^+\pi^0$ amplitude \cite{Blum:2011ng,Blum:2012uk,Blum:2015ywa}, and the first statistically-significant signal of the $\Delta I =1/2$ enhancement~\cite{Boyle:2012ys}, both carried out by RBC-UKQCD collaboration and in agreement with previous analytical computations~\cite{Pich:1995qp,Pich:1990mw,Jamin:1994sv,Bertolini:1997ir,Antonelli:1995nv,Antonelli:1995gw,Hambye:1998sma,Bardeen:1986vz,Buras:2014maa,Bijnens:1998ee}. 

Based on these results, the RBC-UKQCD collaboration reported as well the prediction for the CP violating ratio, $\mathrm{Re}\,(\varepsilon'/\varepsilon) = (1.4 \pm 6.9) \cdot 10^{-4}$ \cite{Bai:2015nea,Blum:2015ywa}, with a tension of $\sim 2\,\sigma$ with Eq.~\eqn{eq:exp}. This result revived again the old debate~\cite{Gisbert:2018tuf,Gisbert:2018niu,Gisbert:2019uyx,Buras:2018ozh}, bringing back the past approaches with low $\varepsilon'/\varepsilon$ predictions~\cite{Buras:2015xba,Buras:2016fys,Buras:2015yba}. However, the same lattice study~\cite{Bai:2015nea} reported a $I=0$ phase shift about $\sim 3 \,\sigma$ away from its dispersive prediction, exhibiting that the long-distance re-scattering of the final pions in $K\to 2\,\pi$ was not well-controlled in the simulation. 

Although  the importance of the pion dynamics in the determination of $\varepsilon'/\varepsilon$ was already pointed out in Refs.~\cite{Pallante:1999qf,Pallante:2000hk,Pallante:2001he}, the results required a thorough revision with an update of the different inputs. At the end of 2017, the updated prediction of $\varepsilon'/\varepsilon$~\cite{Gisbert:2017vvj} was found again in good agreement with Eq.~\eqn{eq:exp}.  

Last year, the authors of Ref.~\cite{Cirigliano:2019cpi} performed the update of the last missing ingredient, isospin breaking corrections to $\varepsilon'/\varepsilon$. Including all the sources of uncertainty, the prediction for the CP violating ratio was found in perfect agreement with Eq.~\eqn{eq:exp}. 

The results of Refs.~\cite{Cirigliano:2019cpi} were presented to the high energy physics community in several international conferences~\cite{Cirigliano:2019obu,Cirigliano:2019zjv,Cirigliano:2019ani}, making necessary a revision of the lattice QCD prediction~\cite{Bai:2015nea,Blum:2015ywa} to confirm the findings of Refs.~\cite{Gisbert:2017vvj,Cirigliano:2019cpi}.

After years of efforts to develop the lattice understanding of the pion dynamics, the RBC-UKQCD outcomes~\cite{Abbott:2020hxn} emerged in April 2020, which are outlined in the following~\cite{Blum:2015ywa,Abbott:2020hxn}:
\begin{itemize}
    \item Lattice strong phase shifts in accordance with dispersive predictions.
    \item $K \to 2\,\pi$ isospin amplitudes $A_{0,2}$ in good agreement with their experimental measurements.
    \item $\varepsilon'/\varepsilon$ prediction in agreement with Eq.~\eqn{eq:exp}~,
    \begin{equation}\label{eq:latt}
\hspace{0.5cm}\mathrm{Re}\, (\varepsilon'/\varepsilon)_{\,\mathrm{lattice}}\; =\;
(22 \pm 8) \cdot  10^{-4}\, ,
\end{equation}
    where isospin breaking effects are not included in the central value.
\end{itemize}

These lattice results close the long debate on the SM prediction of $\varepsilon'/\varepsilon$, confirming the observations made 20 years earlier in $\chi$PT~\cite{Pallante:1999qf,Pallante:2000hk,Pallante:2001he,Cirigliano:2003nn,Cirigliano:2003gt}~, and offering good evidence for their updates~\cite{Gisbert:2017vvj,Cirigliano:2019cpi}.\footnote{Using the recent lattice results~\cite{Blum:2015ywa,Abbott:2020hxn}, the authors of Ref.~\cite{Aebischer:2020jto} have also reported a prediction of $\varepsilon'/\varepsilon$ in agreement with Eq.~\eqn{eq:exp}.}

In the following sections, we highlight the essential dynamical features of $K\to 2\,\pi$ decays and briefly summarize the updated $\varepsilon'/\varepsilon$ prediction from Ref.~\cite{Cirigliano:2019cpi}.

\section{Dynamics of $K \to 2\,\pi$ decays}
\label{sec:anatomy}

The physical kaon decay amplitudes can be expressed in terms of isospin components as 
\begin{eqnarray}  
A(K^0 \to \pi^+ \pi^-) &=&  
\cA_{1/2} + {1 \over \sqrt{2}} \left( \cA_{3/2} + \cA_{5/2} \right)  \, ,
\no\\[2pt]
A(K^0 \to \pi^0 \pi^0) &=& 
\cA_{1/2} - \sqrt{2} \left( \cA_{3/2} + \cA_{5/2}  \right) \, ,\no
\\[2pt]
A(K^+ \to \pi^+ \pi^0) &=&
{3 \over 2}  \left( \cA_{3/2} - {2 \over 3} \cA_{5/2} \right)\,,
\label{eq:2pipar}
\end{eqnarray}
with $\cA_{1/2}= A_{0}\, \mathrm{e}^{i \chi_0}$, $\cA_{3/2}= \frac{2}{5}\,A_{2} \,\mathrm{e}^{i \chi_2}+\frac{3}{5}\,A_{2}^+ \,\mathrm{e}^{i \chi_2^+}$, and $\cA_{5/2}= \frac{3}{5}\,A_{2}\, \mathrm{e}^{i \chi_2}-\frac{3}{5}\,A_{2}^+\, \mathrm{e}^{i \chi_2^+}$. $A_0$, $A_2$ and $A_2^+$ are real and positive amplitudes if the CP-conserving limit is assumed. In the isospin limit, $\cA_{5/2}$ is zero, resulting in only two isospin amplitudes $A_{0}$ and $A_2=A_2^+$ corresponding to final states $(\pi\pi)_{I=0,2}$, in addition to its respective phase differences $\chi_0$ and $\chi_2=\chi_2^+$ which represent the S-wave scattering phase shifts. 

Information regarding amplitudes and phase differences can be extracted from the experimental $K\to 2\,\pi$ branching ratios~\cite{Antonelli:2010yf}:
\begin{eqnarray}
A_0 &=& (2.704 \pm 0.001) \cdot 10^{-7} \mbox{ GeV}, \nn
A_2 &=& (1.210 \pm 0.002) \cdot 10^{-8} \mbox{ GeV}, \nn
\chi_0 - \chi_2 &=& (47.5 \pm 0.9)^{\circ},
\label{eq:isoamps}
\end{eqnarray}
which allow us to learn two things about the dynamics of these decays:
\begin{enumerate}
\item Substantial enhancement in the isoscalar amplitude relative to the isotensor one, 
\be
\hspace{1.cm}\omega \,\equiv\, 
\mathrm{Re}\, A_2 / \mathrm{Re}\, A_0\,\approx\, 1/22~.
\ee
\item Large difference between $\chi_0$ and $\chi_2$, which implies by Eq.~\eqn{eq:2pipar} that half of the ratio $\cA_{1/2}/\cA_{3/2}$ is generated from its absorptive contribution:
\be\label{eq:AbsRat}
\mathrm{Abs}\, (\cA_{1/2}/\cA_{3/2})\approx \mathrm{Dis}\, (\cA_{1/2}/\cA_{3/2})~.
\ee
\end{enumerate}
The above statements are general, and represent excellent control tests for $K\to 2\,\pi$ amplitude predictions when a specific theoretical framework is adopted.

In the presence of CP violation, $\mathrm{Im}\, A_{0,2}$ are nonzero and allows us to define the CP violating observable,\footnote{Eq.~\eqn{eq:cp1} is leading order (LO) in CP violation. Corrections of $\mathcal{O}\left((\mathrm{Im}\,A_{0,2})^2\right)$ are very suppressed.}
\begin{equation}
\varepsilon'\,  =\, 
- \frac{i}{\sqrt{2}} \: e^{i ( \chi_2 - \chi_0 )} \:\omega\;\frac{\mathrm{Im}\, A_{0}}{ \mathrm{Re}\, A_{0}}\,
\left[
1 \, - \,\frac{1}{\omega}
\frac{\mathrm{Im}\, A_{2}}{ \mathrm{Im}\, A_{0}} \right] .
\label{eq:cp1}
\end{equation}
Eq.~\eqn{eq:cp1} shows that $\varepsilon'/\varepsilon$ is approximately real ($\chi_2 - \chi_0 - \pi/2\approx 0$), and also that $\varepsilon'$ is suppressed by the ratio $\omega$. In view of the later, one could be worried about an additional suppression from a possible subtle numerical cancellation emerging from 
\begin{equation}\label{eq:xfirst}
    \hspace{2.cm}x\equiv 1 \, - \,\frac{1}{\omega}
\frac{\mathrm{Im}\, A_{2}}{ \mathrm{Im}\, A_{0}}~.
\end{equation}
The study of Eq.~\eqn{eq:xfirst} requires the estimation of $\mathrm{Im}\, A_{0,2}$, which can be done in the limit of a large number of QCD colours where the T-product of two colour singlet currents factorizes
\begin{equation}
  \hspace{1.5cm}\langle Q_{i}\rangle=\langle J\cdot J\rangle\:=\:\langle J \rangle\:\langle J \rangle\:B_i~, 
\label{eq:factorization}
\end{equation}
and the current $\langle J\rangle$ has a well-known representation at LO in the $1/N_C$ expansion. Here, $B_i$ parametrizes the deviation of the true hadronic matrix element $\langle Q_{i}\rangle$ from its large-$N_C$ approximation $\langle J \rangle\:\langle J \rangle$.

Due to the chiral enhancement of $(V-A)\,\times\,(V+A)$ operators, as well as the size of their respective Wilson coefficients, $\mathrm{Im}\, A_{0,2}$ are mainly dominated by $Q_{6,8}$~, respectively. Considering only these two operators, one finds
\beqn
    x\approx 1+\frac{\omega^{-1}}{2\sqrt{2}}\left(\frac{F_K}{F_\pi}-1\right)^{-1}\frac{y_8(\mu)}{y_6(\mu)}\,\widetilde{B}\,
    \approx\, 1-\frac{1}{2}\,\widetilde{B}~,
    \label{eq:xapprox}
\eeqn
with $\widetilde{B}\equiv\frac{B_8^{(3/2)}}{B_6^{(1/2)}}$. $F_K$ and $F_\pi$ are the kaon and pion decay constants, respectively. The CP-odd components of the Wilson coefficients at the short-distance scale $\mu$ are denoted by $y_i(\mu)$~. Since the anomalous dimensions of $y_6$ and $y_8$ are the same at $N_C\to \infty$, the ratio $y_8(\mu)/y_6(\mu)$ is independent of $\mu$ in that limit, showing how the product of two colour-singlet quark currents factorizes, as previously seen in Eq.~\eqn{eq:factorization}.

In the large-$N_C$ limit, $x\approx 1/2$ and therefore $\mathcal{O}(1/N_C)$ corrections are numerically relevant in Eq.~\eqn{eq:xapprox} . Performing an expansion of $B_6^{(1/2)}$ and $B_8^{(3/2)}$ in powers of $1/N_C$, that is $B_i= 1+\frac{1}{N_C}b_i+\mathcal{O}\left(1/N_C^2\right)$~, where $b_i$ are $\mathcal{O}(1)$~, Eq.~\eqn{eq:xapprox} becomes
\beqn
    \hspace{2.cm}x\approx\frac{1}{2}-\frac{1}{3}(b_8-b_6)~,
    \label{eq:xapprox1}
\eeqn
where $N_C=3$ has been employed. Eq.~\eqn{eq:xapprox1} shows explicitly the possibility of a strong cancellation between the two isospin contributions, reaching is maximal efficiency for $b_8-b_6\approx \frac{3}{2}$, however this cancellation depends on the predicted values of $b_6$ and $b_8$ which are $\mathcal{O}(1)$ as already mentioned.

Although isospin breaking effects are suppressed in almost all the phenomenological observables, $\varepsilon'$ represents an exception. In the second term of Eq.~\eqn{eq:cp1}, we observe that a tiny isospin violating correction in $\mathrm{Im}\, A_{2}$ can break its own suppression through the large enhancement of $\omega^{-1}\approx 22$ consequence of the $\Delta I=1/2$ rule. These effects can be naively accounted in Eq.~\eqn{eq:xapprox1}, inserting an additional parameter $\Omega_{\mathrm{naive}}$~, 
\beqn
    \hspace{1.cm}x\approx\frac{1}{2}-\frac{1}{3}\,(b_8-b_6)+\Omega_{\mathrm{naive}}~.\label{eq:naive}
\eeqn
As an example, we observe that LO isospin breaking corrections originated from $\pi^0$-$\eta$ mixing can be estimated as $\Omega_{\mathrm{naive}}\sim \mathcal{O}\left(\omega^{-1}\,\varepsilon^{(2)}\right)\sim\mathcal{O}(10^{-1})$~, where $\varepsilon^{(2)}\sim \mathcal{O}(10^{-2})$ is the tree-level mixing angle. Therefore, isospin corrections in $x$ are relevant and even more if Eq.~\eqn{eq:xapprox1} results to have a strong cancellation.  

For a proper control of the $\varepsilon'/\varepsilon$ prediction, it is useful to rewrite Eq.~\eqn{eq:cp1} to first order in isospin breaking corrections~\cite{Cirigliano:2003nn,Cirigliano:2003gt}
\be\label{eq:epsp_simp}
\mbox{}\hskip -.6cm
\mathrm{Re}\Bigl(\frac{\varepsilon'}{\varepsilon}\Bigr) \, = \, - \frac{\omega_+}{\sqrt{2}\, |\varepsilon|}  \, \left[
\frac{\mathrm{Im}\, A_{0}^{(0)} }{ \mathrm{Re}\, A_{0}^{(0)} }\,
\left( 1 - \Omega_{\rm eff} \right) - \frac{\mathrm{Im}\, A_{2}^{\rm emp}}{ \mathrm{Re}
  A_{2}^{(0)} } \right]  . \;\;
\ee
The superscript $(0)$ denotes the isospin limit, $\mathrm{Im}\, A_2^{\rm emp}$ contains the $I=2$ contribution from the electromagnetic penguin operator, and $\omega_+ \equiv \mathrm{Re}\, A_{2}^{+}/\mathrm{Re}\, A_{0}$~. The parameter $\Omega_{\rm eff}$ encodes isospin-breaking corrections.

To minimize the theoretical uncertainty in our prediction, the CP-conserving amplitudes $\mathrm{Re}\, A_{0,2}^{(0)}$~, and consequently $\omega_+$, are set to their experimental values, provided in Eq.~(\ref{eq:isoamps}). Only $\mathrm{Im}\, A_{0}^{(0)}$, $\mathrm{Im}\, A_{2}^{(0)}$ and $\Omega_{\rm eff}$ (naively related with $b_6$~, $b_8$ and $\Omega_{\mathrm{naive}}$, respectively) require an analytical computation within a theoretical framework. The later has to account for the dynamics of $K\to 2\,\pi$ decays, raised at the beginning of this section, as well as good theoretical control of the different contributions, to disentangle the possibility of cancellation between different contributions as seen in Eq.~\eqn{eq:naive}. The theoretical framework adopted for our prediction is presented in the following section.

\section{Theoretical framework}

The physical origin of $\varepsilon'/\varepsilon$ is at the electroweak scale where all the flavor-changing processes are defined in terms of quarks and gauge bosons. The gluonic corrections to the $K\to2\,\pi$ amplitudes are amplified with large logarithms due to the existence of very different mass scales ($M_\pi < M_K
\ll M_W$), which can be summed up using the Operator Product Expansion (OPE) and the Renormalization Group Equations (RGEs), all the way down to $\mu<m_c$ scales. Finally, in the three-flavor theory, one gets the following effective Lagrangian~\cite{Buchalla:1995vs} 
\be\label{eq:Leff}
\cL_{\mathrm{eff}}^{\Delta S=1}\, =\, - \frac{G_F}{\sqrt{2}}\,
 V_{ud}^{\phantom{*}}V^*_{us}\,  \sum_{i=1}^{10}
 C_i(\mu) \, Q_i (\mu)\, ,
\ee
which is a sum of local operators weighted by $C_i(\mu)$ short-distance
coefficients which depend on the parameters of the heavy masses ($\mu>M$) and the Cabibbo-Kobayashi-Maskawa (CKM) matrix.
At NLO, the Wilson coefficients $C_i(\mu)$ are known~\cite{Buras:1991jm,Buras:1992tc,Buras:1992zv,Ciuchini:1993vr}. This includes both $\cO(\alpha_s^n t^n)$ and $\cO(\alpha_s^{n+1} t^n)$ corrections, where $t\equiv\log{(M_1/M_2)}$ applies to any ratio logarithm with $M_1,M_2\geq\mu$~. Some next-to-next-to-leading-order (NNLO) corrections are already established \cite{Buras:1999st,Gorbahn:2004my} and an attempt for the full short-distance calculation at the NNLO is underway~\cite{Cerda-Sevilla:2016yzo}.  

We can use symmetry considerations below the resonance region, where perturbation theory no longer works, to describe another effective field theory in terms of the QCD Goldstone bosons ($\pi$, $K$, $\eta$). The pseudoscalar octet dynamics are defined by $\chi$PT through a perturbative expansion in powers of momenta and quark masses over the chiral symmetry breaking scale $\Lambda_\chi\sim 1$~GeV. Then, we can build the effective bosonic Lagrangian with the same SU$(3)_L\otimes \:$SU$(3)_R$ transformation properties as Eq.~\eqn{eq:Leff}, leading to $\cL^{\Delta S=1}=\sum_{n=2}^{n_{\mathrm{cut}}}\cL^{\Delta S=1}_n$, where the cut in $n$ depends on how accurate we want to be in our predictions. At LO, only three terms are allowed by symmetries
\begin{eqnarray}\label{eq:LchiPT}
\cL_{2}^{\Delta S=1}\:=\:G_8\:\cL_{8}\:+\:G_{27}\:\cL_{27}\:+\:G_8\:g_{\mathrm{ewk}}\:\cL_{\mathrm{ewk}}\, .
\end{eqnarray}
The chiral low-energy constants (LECs), {\it i.e.} $G_8$, $G_{27}$ and $G_8\, g_{\mathrm{ewk}}$, can not be determined by symmetry principles, however Eq.~\eqn{eq:factorization} allows us to perform a matching between Eqs.~(\ref{eq:Leff}) and (\ref{eq:LchiPT}) in the large-$N_C$ limit, which gives rise to chiral LECs in terms of Wilson coefficients, and so the short-distance dynamics. Since the large-$N_C$ limit is only applied in the matching, missing $1/N_C$ corrections are not enhanced by large logarithms, {\it i.e.} the naive estimation of these contributions is $b_{6,8}^{\mathrm{matching}}\sim\mathrm{log}(\mu/M_\rho)\sim\mathcal{O}(10^{-1})$, one order of magnitude smaller than its natural size, and then numerically irrelevant to play a game in the cancellation of Eq.~\eqn{eq:naive}. In contrast, the decay amplitudes get large logarithmic corrections from pion loops which are a direct consequence of the large phase shift difference between $\chi_0$ and $\chi_2$. The naive estimation of these contributions implies $b_{6,8}\sim \mathrm{log}(\mu/M_\pi)\sim\mathcal{O}(1)$~, and then they are crucial in Eq.~\eqn{eq:naive}. These corrections can be rigorously calculated using the usual $\chi$PT methods~\cite{Pallante:1999qf,Pallante:2000hk,Pallante:2001he}, and they turn out to be positive for $\left. A_{0}\right|_{Q_6}$, while negative for $\left. A_{2}\right|_{Q_8}$, or in other words $b_8<0$ and $b_6>0$. Therefore, the numerical cancellation between the $I=0$ and $I=2$ terms in Eq.~\eqn{eq:naive} is completely destroyed by the chiral loop corrections, leading to a sizeable enhancement of the SM prediction for $\varepsilon'/\varepsilon$, presented in the next section.

\section{\boldmath SM prediction of $\varepsilon'/\varepsilon$}

Taking into account all computed corrections in Eq.~\eqn{eq:epsp_simp}, which we refer for details to~\cite{Gisbert:2017vvj,Cirigliano:2019cpi}, our SM prediction for $\varepsilon'/\varepsilon$ is
\beqn\label{eq:finalRes}
\hspace{-0.7cm}\mbox{Re}\left(\varepsilon'/\varepsilon\right)&\!\!\! =&\!\!\!\left(13.8\pm 1.3_{\,\gamma_5}\pm 2.5_{\,\mathrm{LECs}}\pm 3.5_{1/N_C}\right)\cdot 10^{-4}
\no\\[5pt] &\!\!\! =&\!\!\! 
\left(14\pm 5\right) \cdot 10^{-4}\, .
\eeqn
where isospin breaking corrections have been included, and result in $\Omega_{\rm eff}=0.11\pm 0.09$~\cite{Cirigliano:2019cpi}. The first error represents the uncertainty from NNLO corrections to the Wilson coefficients accounted through the choice of the scheme for $\gamma_5$.  The second error comes from the input values of the strong LECs (mainly $L_{5,7,8}$).  The last error parametrizes our ignorance about $1/N_C$-suppressed contributions in the matching  region  which have been estimated conservatively through the variation of $\mu$ and $\nu_\chi$ in the intervals $[0.9,1.2]$~GeV and $[0.6,1]$~GeV, respectively. 

Our SM prediction for $\varepsilon'/\varepsilon$ is in perfect agreement with the measured experimental value. We have shown the important role of FSI in $K^0\to\pi\pi$. When $\pi\pi$ re-scattering corrections are taken into account, the numerical cancellation between the $Q_6$ and $Q_8$ terms in Eq.~\eqn{eq:naive} is no longer operative because of the positive enhancement of the $Q_6$ amplitude and the negative suppression of the $Q_8$ contribution. 

Although our theoretical uncertainty is still large, improvements can be achieved in the next years:
\begin{itemize}
\item NNLO computation of the Wilson coefficients is close to being finished~\cite{Cerda-Sevilla:2016yzo}. 
\item The $\cO(e^2p^0)$ coupling $G_8\, g_{\mathrm{ewk}}$ can be expressed as a dispersive integral over the hadronic vector and axial-vector spectral functions. The $\tau$ decay data can be used to perform a direct determination of this LEC~\cite{Sanchez:2018atj,AntonioS}.

\item  Estimation of higher-order chiral logarithmic corrections could be feasible either through explicit two-loop calculations or with dispersive techniques~\cite{Pallante:1999qf,Pallante:2000hk,Pallante:2001he,Buchler:2001nm,Buchler:2005xn}.

\item A matching calculation of the weak LECs at NLO in $1/N_C$ remains a very challenging task. A fresh view to previous attempts \cite{Bertolini:1995tp,Bertolini:1997nf,Bertolini:1998vd,Bijnens:2000im,Hambye:2003cy,
Pich:1995qp,Pich:1990mw,Jamin:1994sv,Bertolini:1997ir,Antonelli:1995nv,Antonelli:1995gw,Hambye:1998sma,Bardeen:1986vz,Buras:2014maa,Bijnens:1998ee} could suggest new ways to face this unsolved problem.

\end{itemize}

\section*{Acknowledgements}

I would like to thank Vincenzo Cirigliano, Toni Pich, and Antonio Rodríguez for enjoyable collaborations. This work is supported by the \textit{Bundesministerium für Bildung und Forschung – BMBF}.

\end{document}